\documentclass{webofc}
\usepackage[varg]{txfonts}   
\usepackage{lineno}
\newcommand{\sqrtsNN}{\mbox{$\sqrt{\mathrm{s}_{_{\mathrm{NN}}}}$}}
\begin{document}
\title{QCD Critical Point and Net-Proton Number Fluctuations at RHIC-STAR}
%
%

\author{\firstname{Yu} \lastname{Zhang} \inst{1,2}\fnsep\thanks{\email{zhang_yu@mails.ccnu.edu.cn}} 
    \\ for the STAR Collaboration
}

\institute{Central China Normal University 
\and
           Lawrence Berkeley National Laboratory 
}

\abstract{%
  In the search of QCD phase boundary and critical point, higher-order cumulants of conserved quantities are proposed as promising observables and have been studied extensively both experimentally and theoretically. In this paper we present cumulant ratios up to $6^{th}$-order of net-proton number distributions in Au+Au collisions at \sqrtsNN = 7.7 - 200 GeV from STAR Beam Energy Scan program phase I and $\sqrt{\mathrm{s}}$ = 200 GeV $p+p$ collisions. The results are compared with various models and Lattice QCD calculations.}
\maketitle

\section{Introduction}
In heavy-ion collision physics it is predicted that under a very high temperature and baryon density a deconfined quark-gluon plasma (QGP) phase can be created and studying the QCD phase structure is one of the main goals. A phase diagram in terms of temperature and baryon chemical potential ($\mu_{\mathrm{B}}$) is usually used to explore the QCD phase structure. Regarding the phase transition between the QGP phase and hadronic phase, first principle Lattice QCD calculation~\cite{nature_Aoki} at $\mu_{\mathrm{B}}$ = 0 MeV suggests a smooth crossover transition. While at large $\mu_{\mathrm{B}}$, various QCD-based models predict first order phase transition~\cite{1st_1}. Thermodynamically there should be an end point of the first order phase boundary which is called QCD critical point. The possible QCD critical point and the first order phase boundary have been investigated both experimentally~\cite{PhysRevLett.112.032302,PhysRevLett.113.092301,PhysLettB785.551} and theoretically.

Higher-order cumulants of conserved quantities like net-baryon ($B$), net-charge ($Q$) and net-strangeness ($S$) are proposed as promising observables to search for the QCD critical point and the first order phase boundary. Higher-order cumulants are sensitive to the correlation length ($\xi$)~\cite{PhysRevLett.103.262301} and are directly related to susceptibility ($\chi$) of the system~\cite{susceptibility}. It is predicted that the fourth-order fluctuations will exhibit a non-monotonic energy dependence~\cite{PhysRevLett.107.052301,PhysRevD.85.034027,PhysRevD.95.014038} when passing through the critical region. For $5^{th}$- and $6^{th}$-order cumulants recent calculations from Lattice QCD~\cite{PhysRevD.101.074502} and the functional renormalisation group approach (FRG)~\cite{fu2021hyperorder} show that they will be negative due to the crossover transition between QGP and hadronic phase. 
At high baryon density region, on the other hand, they are also sensitive to the first order phase boundary~\cite{volker2}.    
\section{Data and Analysis}
The data of Au+Au collisions at $\sqrtsNN$ = 7.7 - 200 GeV and $p+p$ collisions at $\sqrt{\mathrm{s}}$ = 200 GeV are collected in RHIC Beam Energy Scan program phase I. Protons and antiprotons are identified by the Time Projection Chamber (TPC) and Time of flight (TOF) detectors at rapidity window $-0.5<y<0.5$ and transverse momentum window $0.4<p_{\mathrm{T}}<2.0$ GeV/$c$.

The centrality is determined using charged particle multiplicity within $|\eta|<1.0$ excluding protons and antiprotons to avoid auto-correlation effect~\cite{Luo_2013}. The centrality bin width correction~\cite{Luo_2013} is applied to suppress initial volume fluctuation effect. Cumulants are calculated at each multiplicity bin and then their weighted averages are taken for each centrality bin. The weight is the number of events at the corresponding multiplicity bin. Detector efficiency correction~\cite{PhysRevC.91.034907} in cumulant calculations are done by assuming binomial detector efficiency. Statistical uncertainties of cumulants are estimated by Bootstrap~\cite{bootstrap} and Delta methods~\cite{Luo_2012}.
%
%
\section{Results}
\begin{figure}[htbp]
	\centering
	\includegraphics[width=0.80\linewidth]{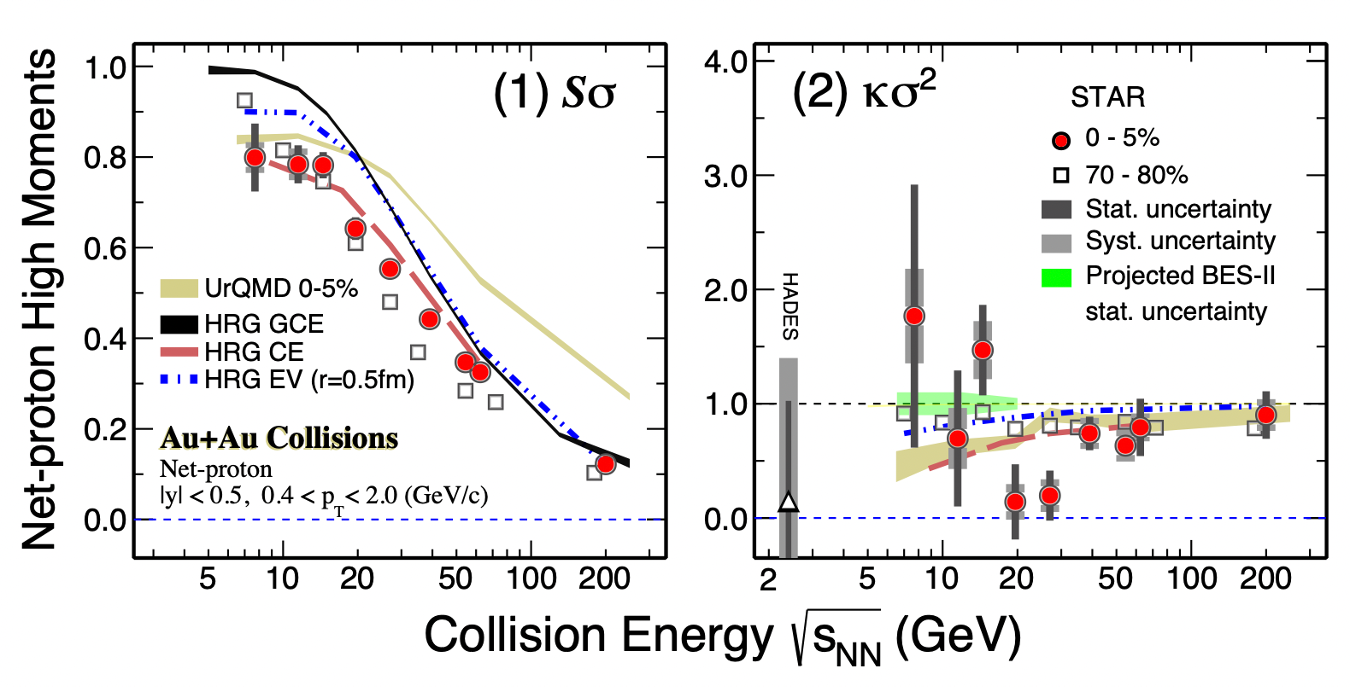}
	\caption{Energy dependence of $S\sigma$ and $\kappa\sigma^2$ of net-proton number distributions in Au+Au collisions at $\sqrt{\mathrm{s_{NN}}}$ = 7.7 - 200 GeV. The calculations from different conditions (GCE, EV, and CE) of hadron resonance gas model (HRG) and the hadronic transport UrQMD model are shown as black, red, blue bands, and a gold band respectively.}
	\label{fig:c4} 
\end{figure}

\begin{figure}[htbp]
	\centering
	\begin{minipage}[t]{0.35\textwidth}
		\centering
		\includegraphics[width=4.5cm]{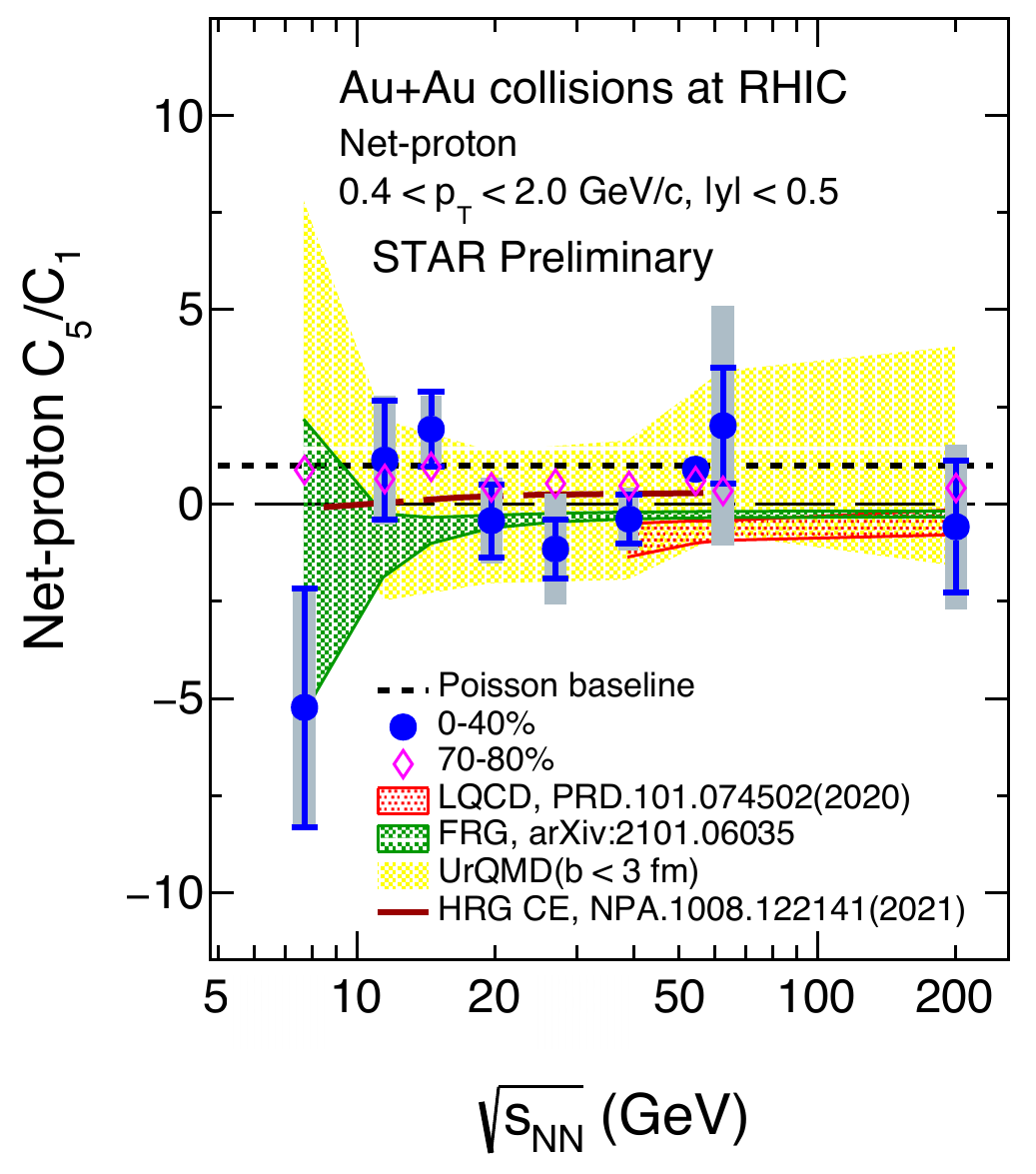}
	\end{minipage}
	\begin{minipage}[t]{0.35\textwidth}
		\centering
		\includegraphics[width=4.5cm]{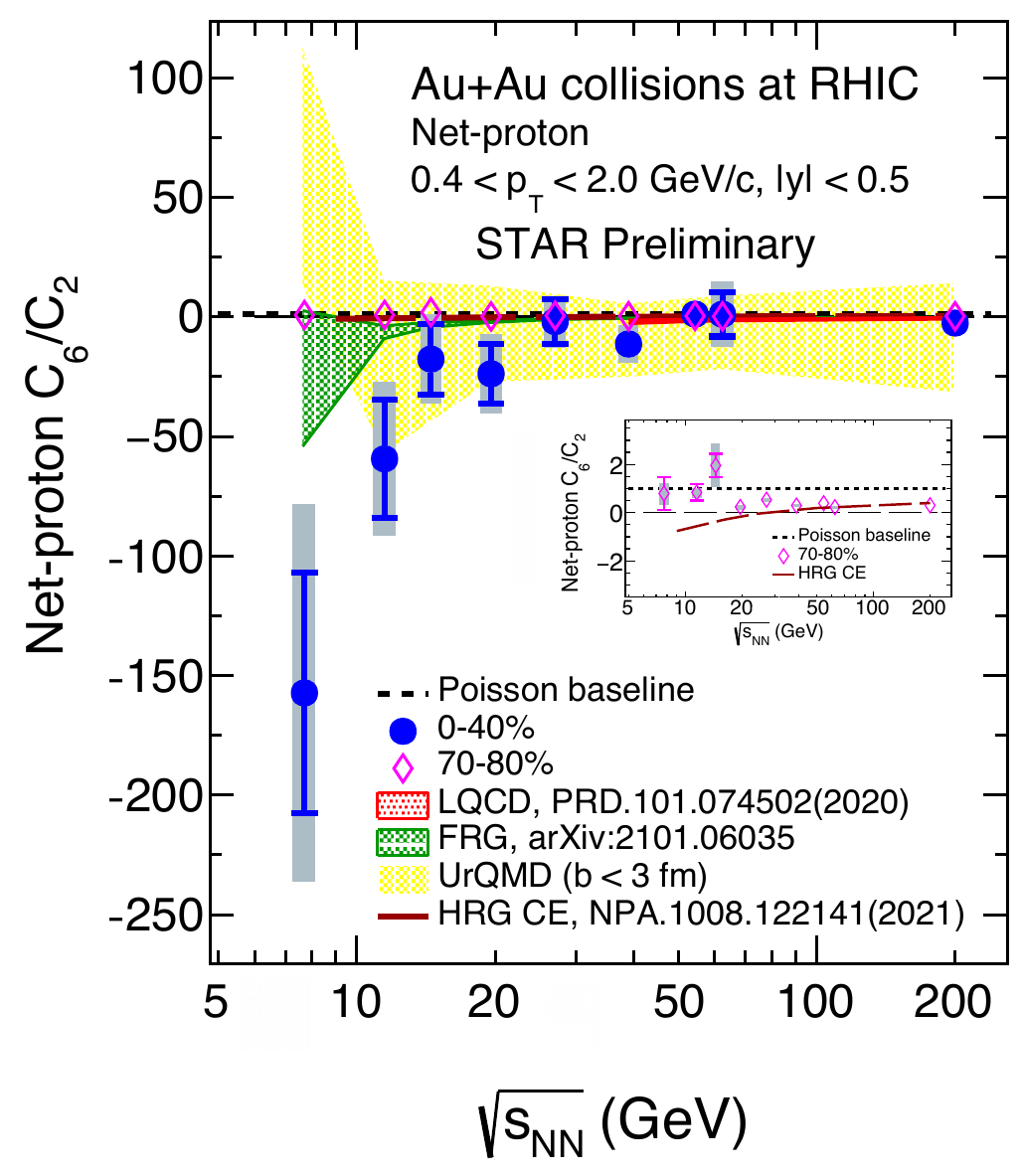}
	\end{minipage}

	\caption{Energy dependence of $\mathrm{C_{5}/C_{1}}$ and $\mathrm{C_{6}/C_{2}}$ of net-proton number distributions in Au+Au collisions at $\sqrtsNN$ = 7.7 - 200 GeV. The calculations from FRG model, Lattice QCD, hadronic transport UrQMD model, and Hadron Resonance Gas (HRG) model are shown as red, green, and yellow bands and red lines respectively.}
	\label{fig:auau}	
\end{figure}
Figure \ref{fig:c4} shows energy dependence of $S\sigma$ and $\kappa\sigma^2$ of net-proton distributions from 0-5\% and 70-80\% centrality bins within $|y|<0.5$ and $0.4<p_{T}<2.0$ GeV/$c$ in Au+Au collisions at $\sqrtsNN$ = 7.7 - 200 GeV~\cite{PhysRevLett.126.092301,starcollaboration2021cumulants}. The $S\sigma$ (left panel) shows a decreasing trend with the increase of collision energy both in central and peripheral collisions. The decreasing trend can be qualitatively described by HRG~\cite{hrg} and UrQMD~\cite{urqmd_model} models. The $\kappa\sigma^{2}$ (right panel) shows a non-monotonic energy dependence in central collisions while the results for peripheral collisions show monotonic energy dependence. The non-monotonic trend in central collisions can not be described by different conditions (GCE, EV, and CE) of HRG and UrQMD models.

Figure \ref{fig:auau} shows energy dependence of $\mathrm{C_{5}/C_{1}}$ and $\mathrm{C_{6}/C_{2}}$ of net-proton distributions from 0-40\% and 70-80\% centrality bins within $|y|<0.5$ and $0.4<p_{T}<2.0$ GeV/$c$ in Au+Au collisions at $\sqrtsNN$ = 7.7 - 200 GeV. It is suggested from Lattice QCD and FRG calculations that $\mathrm{5^{th}}$- and $\mathrm{6^{th}}$-order cumulants show negative signs while calculations from UrQMD and HRG models are consistent with either zero or unity. In UrQMD and HRG models, no phase transition physics is implemented. The measurements of BES-I data are shown as blue circles for 0-40\% and red diamonds for 70-80\%.  
The cumulant ratio $\mathrm{C_{5}/C_{1}}$ (left panel) deviates from zero with less than 2$\sigma$ significance. In peripheral collisions $\mathrm{C_{5}/C_{1}}$ shows positive sign for all energies. The ratio $\mathrm{C_{6}/C_{2}}$ (right panel) for 0-40\% decreases with negative sign with less than 2$\sigma$ significance when decreasing energy while it shows positive sign in peripheral collisions (70-80\%) for all energies.

\begin{figure}[htbp]
	\centering
	\includegraphics[width=0.95\linewidth]{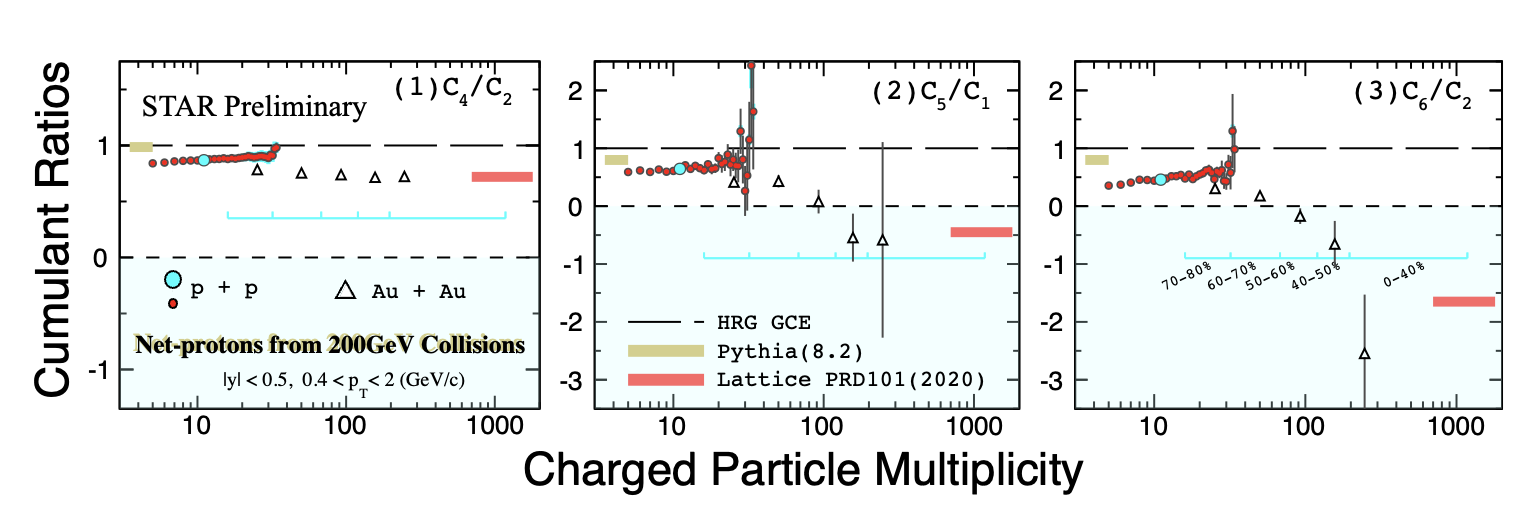}
	\caption{Multiplicity dependence of $\mathrm{C_{4}/C_{2}}$, $\mathrm{C_{5}/C_{1}}$, and $\mathrm{C_{6}/C_{2}}$ of net-proton number distributions in $p+p$ collisions at $\sqrt{\mathrm{s}}$ = 200 GeV. The calculations of HRG model, Pythia (version 8.2), and Lattice QCD are shown as black dashed line, yellow, and red bands respectively.}
	\label{fig:pp} 
\end{figure}
Figure \ref{fig:pp} shows multiplicity dependence of net-proton results of $\mathrm{C_{5}/C_{1}}$ and $\mathrm{C_{6}/C_{2}}$ within $|y|<0.5$ and $0.4<p_{T}<2.0$ GeV/$c$ in $p+p$ collisions at $\sqrt{\mathrm{s}}$ = 200 GeV. We see that the cumulant ratios ($\mathrm{C_{4}/C_{2}}$, $\mathrm{C_{5}/C_{1}}$, and $\mathrm{C_{6}/C_{2}}$) from $p+p$ collisions fit into the multiplicity dependence of results from Au+Au collisions which are shown with triangles. The cumulant ratios $\mathrm{C_{5}/C_{1}}$ and $\mathrm{C_{6}/C_{2}}$ are negative for 0-40\% in Au+Au collisions and positive at peripheral Au+Au collisions and $p+p$ collisions. Pythia~\cite{pythia} calculation using version 8.2 of $p+p$ collisions at $\sqrt{\mathrm{s}}$ = 200 GeV is positive as shown with yellow bands. The Lattice QCD calculation~\cite{PhysRevD.101.074502} at $\sqrt{\mathrm{s_{NN}}}$ = 200 GeV is negative as shown with red bands. Compared with calculations from various models, it is suggested that the negative sign for central Au+Au collisions at $\sqrt{\mathrm{s_{NN}}}$ = 200 GeV is due to a smooth crossover transition between partonic and hadronic phases.

\section{Summary}
In this proceedings, we report the measurements of net-proton cumulant ratios up to $6^{th}$-order in Au+Au collisions at $\sqrtsNN$ = 7.7 - 200 GeV and $p+p$ collisions at $\sqrt{s}$ = 200 GeV from STAR. With results from 200 GeV $p+p$ collisions and the energy dependence of $\mathrm{C_{4}/C_{2}}$, $\mathrm{C_{5}/C_{1}}$, and $\mathrm{C_{6}/C_{2}}$ from the BES-I data sets, and the comparison with LQCD calculations, we conclude: 1) QCD matter is indeed created in the 200 GeV central (0-5\%) Au+Au collisions at RHIC; 2) non-monotonic energy dependence of $\mathrm{C_{4}/C_{2}}$ is observed from the most central (0-5\%) Au+Au collisions~\cite{PhysRevLett.126.092301,starcollaboration2021cumulants}.
Future results from BES-II and STAR fixed-target experiment $\sqrtsNN$ = 3 GeV data sets will allow to answer if QCD critical point exists in the covered energy region. In 2023 to 2025, STAR plans to collect 15 to 20 billion events of Au+Au collisions at $\sqrtsNN$ = 200 GeV. This will allow us to perform more precise measurements of higher-order cumulants, maybe even up to $8^{th}$-order.
 
\section*{Acknowledgments}
This work was supported by the National Key Research and Development Program of China (Grant No. 2020YFE0202002 and 2018YFE0205201), the National Natural Science Foundation of China (Grant No. 11828501, 11890711 and 11861131009) and China scholarship council (No. 201906770055).

\end{document}